%% file: main.tex
\documentclass{llncs}
\usepackage{graphicx}
\usepackage{color}
\usepackage{longtable}
\usepackage{amsfonts}
\usepackage{amsmath}
\usepackage{epsfig}
\usepackage{epstopdf}

\usepackage[utf8]{inputenc}
\begin{document}

\title{CSMA/CA Bottleneck Remediation In Saturation Mode With New Backoff Strategy}
\author{Baher {Mawlawi}$^{1,2,3}$, Jean-Baptiste {Doré}$^{1}$}
     \institute{ CEA-Leti Minatec, 17 rue des Martyrs, 38054 Grenoble Cedex 9,  France
\email{baher.mawlawi@cea.fr} \email{jean-baptiste.dore@cea.fr}
\and University of Lyon, INRIA, Villeurbanne, France
\and INSA-Lyon, CITI-INRIA, F-69621, Villeurbanne, France}
\maketitle

\begin{abstract}
Many modern wireless networks integrate carrier sense multiple
access/collision avoidance (CSMA/CA) with exponential backoff
as medium access control (MAC) technique.
In order to decrease the MAC overhead and the collision
probability, we propose in this paper a new backoff strategy
leading to better saturation throughput and access delay performance
comparing to the classical protocol. We investigate the CSMA/CA
with RTS/CTS technique, and we show that our strategy reaches
better saturation throughput and access delay especially in dense networks. This proposed strategy distributes users over all the backoff stages to solve the bottleneck problem present in the first backoff stage. Finally, we analyze our strategy and we compare it to the classical one modeled by Markov chain.
Analytical and simulation results show the improvment in term of saturation throughput. Cumulative density function (CDF) of the access delay illustrates the important gain obtained by the proposed strategy.
\end{abstract}

\hfill

\section{Introduction}
\input{intro}
\section{CSMA/CA protocol MAC layer}
\label{IEEE 802.11 MAC layer}
\input{standard}

\section{{Proposed Backoff Algorithm Description}}
\label{protocol}
\input{proposed}

\section{Numerical Results}
\label{numerical_results}
\input{numerical_results}

\section{Conclusion}
\label{conclusion}
\input{conclusion}

\section{Acknowledgment}
\input{acknowledgment}

\bibliographystyle{IEEEtran}
\bibliography{bibliography}

\end{document}

%% file: intro.tex
The design of wireless networks for local area communication attracted much of interest \cite{intro1}.
Carrier Sense Multiple Access/Collision Avoidance (CSMA/CA) protocols rely on a decentralized random packet transmission for the efficient use of shared medium. The key features of CSMA/CA is that each link with a pair of transmitter and receiver first senses the medium and transmits a packet only if the channel is sensed idle. Due to its simple and distributed nature, it has been considered as one of the most practical MAC protocols for wireless network. It has been adopted for instance for Wireless Local Area Networks (WLANs) through study group 802.11 \cite{intro3} and for Wireless Sensor Networks (WSNs) through 802.15.4 \cite{802154a}. This family of random access protocols is also a good candidate for future communication systems such as cognitive radio \cite{rawpeach}, Machine to Machine (M2M) wireless networks and so on.

Many previous works try to improve the throughput performance by attempting to optimize the contention window \cite{parallel} \cite{window1} \cite{window2}, but they are based on the IEEE 802.11 backoff strategy.\\
In \cite{novel}, the throughput and the average access delay for different backoff algorithms were studied by simulation only.
In the classical CSMA/CA protocol with 802.11 backoff strategy modeled by Bianchi, it's clearly seen that the first state is the bottleneck of the system, especially in charged mode.
In order to improve the throughput and the system delay we propose and develop in this paper a mathematical model for a new backoff strategy based on Markov chain and we analytically prove that the outcome of the new strategy is better than the classical one in terms of saturation throughput and statistical access delay.

The paper is outlined as follows. We briefly review in Section \ref {IEEE 802.11 MAC layer} the RTS/CTS mechanisms of CSMA/CA protocol. In Section \ref{protocol} we explain the proposed backoff strategy and we give a throughput analytical model. Section \ref{numerical_results} presents the numerical results of the proposed protocol and a comparision with the classical protocol.
Finally, Section \ref{conclusion} is reserved for conclusion.

%% file: standard.tex
In order to describe the CSMA/CA protocol, we propose to explain the well known 802.11 CSMA/CA MAC layer.
The basic medium access mechanism of IEEE 802.11 is DCF (Distributed Coordination Function) which uses CSMA/CA algorithm to serve shared medium access.
It contains two different variants of access method, the basic access method and the optional channel access method with $\textit{request-to-send}$ (RTS) and $\textit{clear-to-send}$ (CTS) exchange.
As the latter variant introduces more efficient performance in term of average throughput \cite{basicpaper} and solves the hidden node problem \cite{hidennodes}, CSMA/CA with RTS/CTS will be investigated in this paper.
Let's consider a network with many terminals and one access point.
If the channel is busy for the transmitters, each one chooses randomly a backoff time (measured in time slots) in the interval [0, $\textit{CW}$) where $CW$ is the $\textit{contention window}$.
As long as the channel is sensed idle for a DIFS, i.e. $\textit{distributed inter-frame space}$ time, the timer (backoff) is decreased by one. When the channel is busy the timer counter is blocked and it resumes when the channel is idle again for at least a DIFS period. $CW$ is an integer between $CW_{min}$ and $CW_{max}$.
After each unsuccessful transmission, $CW$ is doubled up to the maximum value equal to $CW_{max}$+1. The source transmits an RTS frame when the backoff reaches zero and wait for transmission permission (CTS) from the potential receiver before sending the current data packet.
All stations located in the sender’s range that hear the RTS packet should update their NAVs (Network Allocation vector) and defer their transmissions for the duration specified by the RTS. By this strategy, the transmission of data packets and the corresponding ACK can proceed without interference from other nodes.
In addition, whenever erroneous frame is detected by a node, it defers its transmission by a fixed duration indicated by EIFS, i.e., $\textit{extended inter-frame space}$ time.
The contention window is initialized to $CW_{min}$ (minimum contention window).
Dense networks cause collisions between transmitters. Each station involved in the collision double the size of its contention window. In case of a successful transmission, the transmitter re-initialize its contention window by $CW_{min}$.

%% file: proposed.tex
\subsection{Description}

\begin{figure}[tb]
\begin{center}
\includegraphics[width=15pc,keepaspectratio=true]{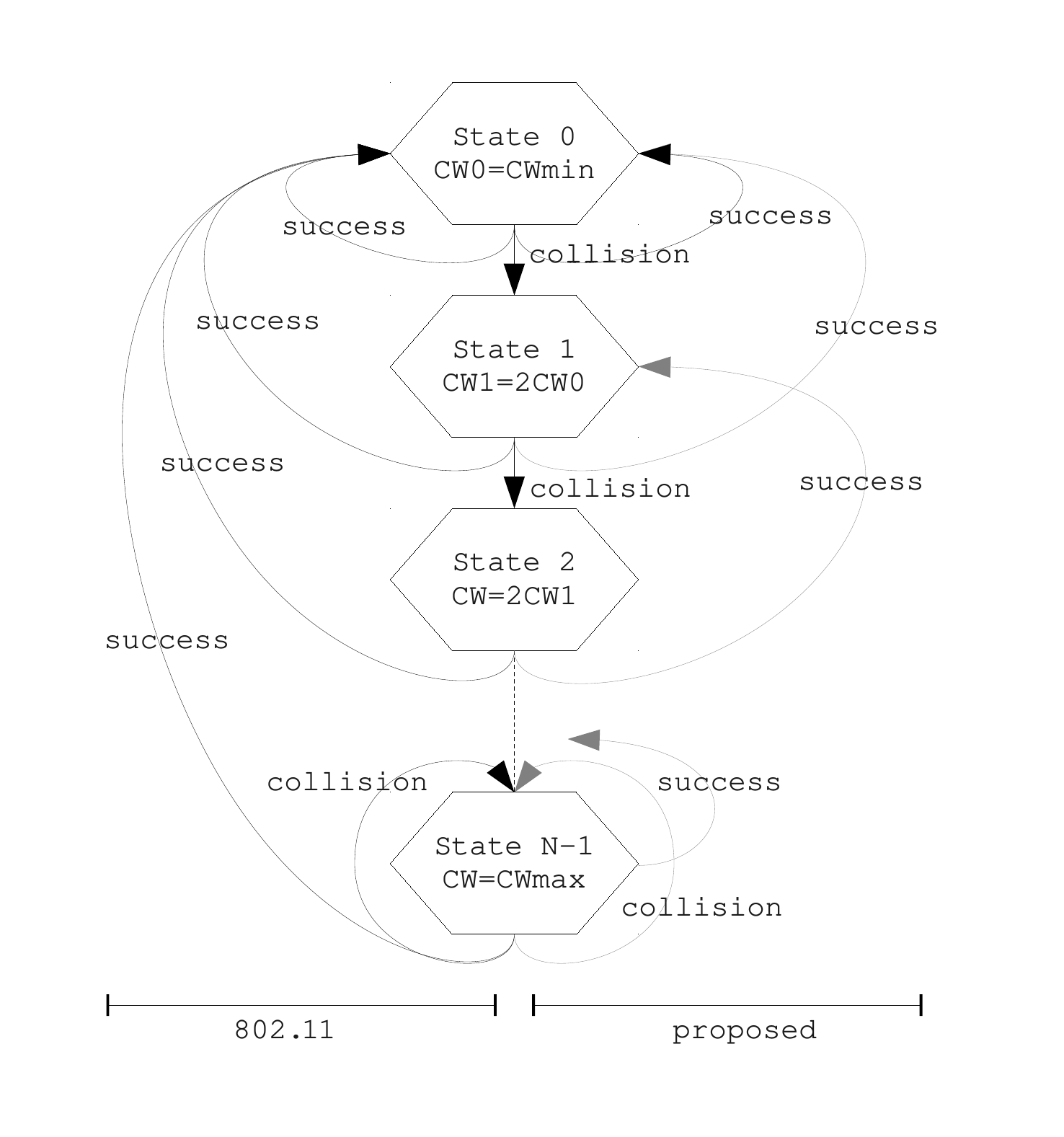}
\caption{802.11 and proposed backoff strategy.}
\label{proposed}
\end{center}
\end{figure}

As explained in Section \ref{IEEE 802.11 MAC layer}, when a station transmits successfully it returns directly to the first backoff stage. This fact introduces a high collision probablity as well as an enormous additive transmission delay due to the high number of users in the same backoff stage ($m=1$). This situation is seen as bottleneck problem.
The proposed CSMA/CA is quite similar to the IEEE 802.11 protocol (Section \ref{IEEE 802.11 MAC layer}), the main difference remains in case of a successful (i.e. collision-free) transmission, the transmitting station reduces the value of its contention window by half, so as to keep its contention window at least  equal to $CW_{min}$ (see Figure \ref{proposed}).

\subsection{System Model}
\begin{figure}[tb]
\begin{center}
	\includegraphics[width=1\columnwidth]{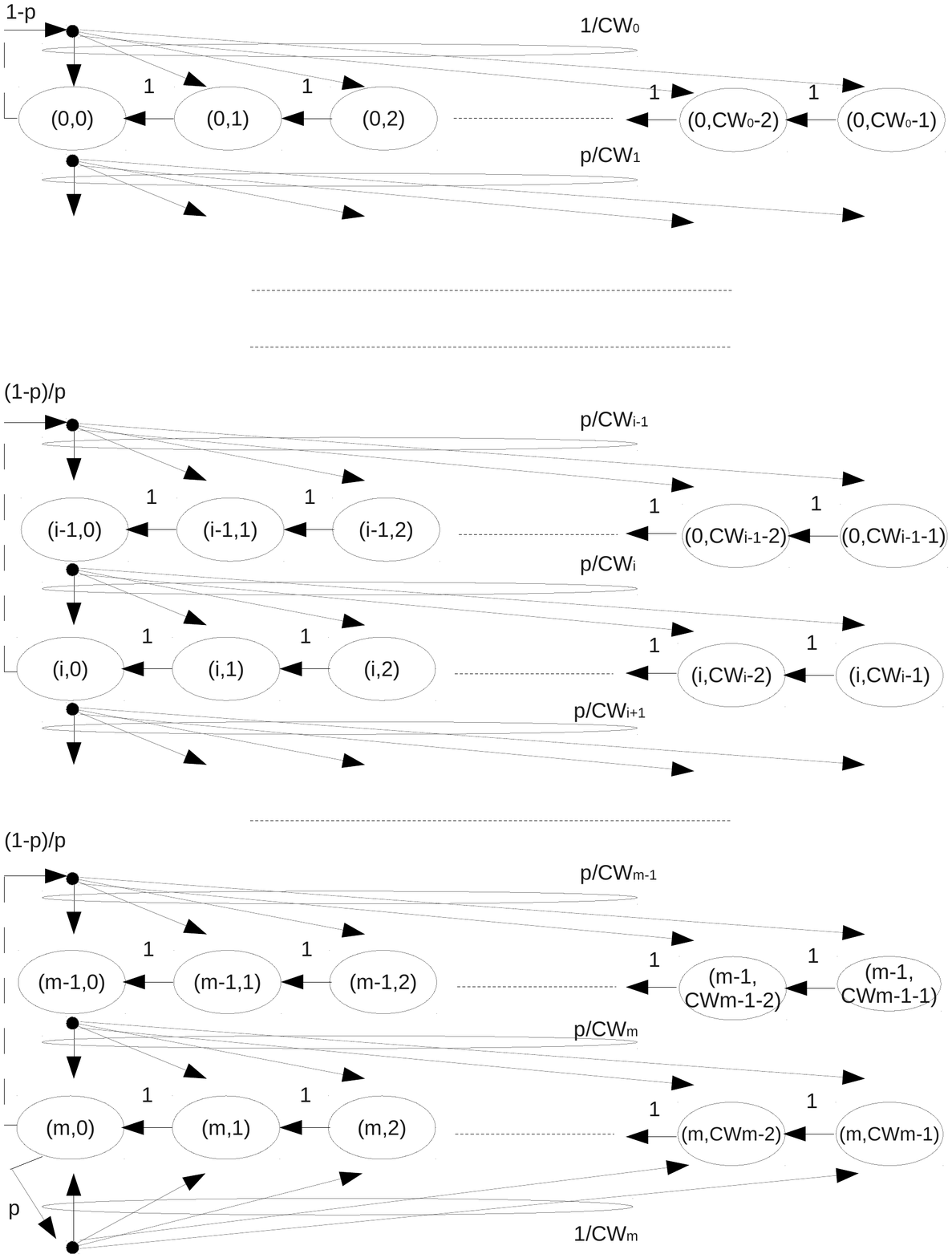}\\
	\caption{Markov chain model of backoff window size in proposed CSMA/CA.}
	\label{markovmodel}
\end{center}
\end{figure}

In this section we will give the model, the analytical calculation of saturation throughput for the proposed CSMA/CA backoff strategy.
We compute the probability of a packet transmission failure due to collision by assuming the following hypothesis \cite{these}:
\begin{itemize}
\item No hidden terminal and capture effect.
\item Failed transmissions only occur as a consequence of collision.
\item All stations are saturated, always having packets to send.
\item For any given station, the probability of collision, $p$, is constant and independent of the station’s collision history of the station and all other stations.
\item The probability of collision does not depend on the backoff stage at which the transmission is made.
\item All users have same bitrates and same amount of time to transmit.
\end{itemize}
Using the Bianchi's model described in \cite{basicpaper} \cite{bianchi1} we model the proposed protocol by a  Markov chain of $m+1$ backoff stages as illustrated Figure \ref{markovmodel}. Each stage of the Markov chain modelled the backoff counter. The number of states per stage is equal to the maximum authorized value of the backoff counter, i.e $CW_{i}$ . It should be mentioned that we use notations described in \cite{basicpaper}, i.e $ CW_{i }=2^i(CW_{min}+1)$.

When a collision occurs a transition from stage $i$ to $(i+1)$ is considered and a random backoff will be chosen between 0 and $CW_{i}$-1 with probability of  $\frac{p}{CW_{i}}$. A successful transmission is modelled by a transition from stage $(i+1)$ to $i$ and a random backoff will be chosen between 0 and $CW_{i-1}$-1 with probability of $\frac{1-p}{CW_{i-1}}$.

Each state of this Markov process is represented by \{$s(t),b(t)$\}, where $b(t)$ is the stochastic process representing the backoff time counter for a given station and $s(t)$ is the stochastic process representing the backoff stage $(0,1,...m)$ of the station at time $t$ \cite{basicpaper}. 
A discrete and integer time scale is adopted where $t$,$(t+1)$ stands for the beginning of two consecutive slot times.

We define $p$ as the probability that, in a slot time, at least one of the $N-1$ remaining stations transmits. This probability can be expressed by: 
\begin{equation}
\label{collision}
p=1-(1-\pi)^{(N-1)}
\end{equation}
Where $\pi$ is the probability that a station transmits a packet. It can be written by:
\begin{equation}
\label{transmission}
\pi=\displaystyle\sum_{i=0}^{m}b_{i,0}
\end{equation}
Where $b_{i,k}$=$\lim\limits_{t \to \infty}P\{s(t)=i,b(t)=k\},i \in (0,m), k \in(0,CW_{i}-1)$ is the stationary distribution of the chain. Only $b(i,0)$ are considered because a transmission occurs when the backoff time counter is equal to zero.
By considering the proposed Markov chain, $b_{i,0}$ can be expressed as a function of $p$:
\begin{equation}
\left\lbrace
\begin{array}{ll}
b_{i,0}=(\frac{p}{1-p})^{i}b_{0,0} & 0<i \leq m  \\ 
b_{i,k}=\frac{CW_{i}-k}{CW_{i}}b_{i,0} & 0<i\leq m, \hspace{1 cm} 0 \leq k\leq CW_{i}-1
\end{array}
\right.
\label{sys}
\end{equation}
It should be noticed that this expression is different from the one expressed in \cite{basicpaper}, due to the proposed backoff strategy. By imposing the classical normalization condition and considering Equation \ref{sys}, $b_{0,0}$ can be expressed as a function of $p$:
\begin{equation}
\begin{split}
1=&\displaystyle\sum_{i=0}^{m}\displaystyle\sum_{k=0}^{CW_{i}-1}b_{i,k}\\
=&\frac{b_{0,0}}{2}\left(W_{min}+1+W_{min}\frac{(1-p)^m-(2p)^m}{(1-3p)(1-p)^{m-1}}\right. \\
 &\left.+\frac{(1-p)^m-p^m}{(1-2p)(1-p)^{m-1}}\right) \\
\end{split}
\label{b00}
\end{equation}
Where $W_{min} = CW_{min}-1$.
Finally, combining equations (\ref{transmission}),(\ref{sys}), and (\ref{b00}), the channel access probability $\pi$ is equal to:
\begin{equation}
\label{access}
\begin{split}
\pi=&\displaystyle\sum_{i=0}^{m}b_{i,0}\\
=&\displaystyle\sum_{i=0}^{m}\left(\frac{p}{1-p}\right)^ib_{0,0}\\
=&b_{0,0}\frac{(1-p)^m-p^m}{(1-2p)(1-p)^{m-1}}\\
\end{split}
\end{equation}
This two equations, (\ref{collision}) and (\ref{access}), form a system of two nonlinear equations that has a unique solution and can be solved numerically for the values of $p$ and $\pi$. 

The saturation throughput, which is the average information payload in a slot time over the average duration of a slot time, can be expressed using the classical expression \cite{basicpaper}:
\begin{equation}
\begin{split}
\tau=&\frac{E[\textit{Payload information transmitted in a slot time}]}{E[\textit{Duration of slot time}]}\\
=&\frac{P_{s}P_{tr}L}{P_{s}P_{tr}T_{s}+P_{tr}(1-P_{s})T_{c}+(1-P_{tr})T_{id}}
\end{split}
\end{equation}
where $P_{tr}=1-(1-\pi)^N$ is the probability that there is at least one transmission in the considered slot time; $L$ is the average packet payload size; $T_{s}$ is the average time needed to transmit a packet of size $L$ (including the inter-frame spacing periods \cite{bianchi1} ); $P_{s}=\frac{N\pi(1-\pi)^{N-1}}{1-(1-\pi)^N}$ is the probability of a successful transmission; $T_{id}$ is the duration of the idle period (a single slot time); and $T_{c}$ is the average time spent in the collision. $T_{c}$ and $T_{s}$ can be calculated for RTS/CTS transmission mode with \cite{basicpaper}:
\begin{equation}
\left\lbrace
\begin{array}{ll}
\begin{split}
T_{s}=&RTS+SIFS+\sigma+CTS+SIFS+\sigma+H+L 
\\ +&SIFS+\sigma +ACK+DIFS+\sigma  \\ 
T_{c}=&RTS+DIFS+\sigma
\end{split}
\end{array}
\right.
\end{equation}
where $H$, $L$, and $ACK$ are the transmission times needed to send the packet header, the payload, and the acknowledgment, respectively. $\sigma$ is the propagation delay.

%% file: numerical_results.tex
In this Section we study the validity, the saturation throughput and the delay of the analytical proposed model. The system of two nonlinear equations (\ref{collision}) and (\ref{access}) is solved numerically. The protocol and channel parameters adopted are those specified in Table \ref{parameters}. However analysis and results can be extended to others PHY layers.  The minimal contention window ($W_{min}$) has been chosen constant and equal to 16.
\begin{table}[ht!]
\caption{PHY layer parameters}
\centering
\begin{tabular}{lcc|}
\hline
Packet payload & 8184 bits\\
MAC header & 272 bits \\
PHY header & 128 bits \\
ACK length & 112 bits + PHY header \\
RTS length & 160 bits + PHY header \\
CTS length & 112 bits + PHY header \\
Channel Bit Rate & 1 Mbit/s  \\
Propagation Delay & 1 $\mu$s \\
SIFS & 28 $\mu$s \\
Slot Time & 50 $\mu$s\\
DIFS & 128 $\mu$s\\
\hline
\end{tabular}
\label{parameters}
\end{table}
\subsection{Validation Of Analytical Results}
\begin{figure}[h!]
\begin{center}
	\includegraphics[width=1\columnwidth]{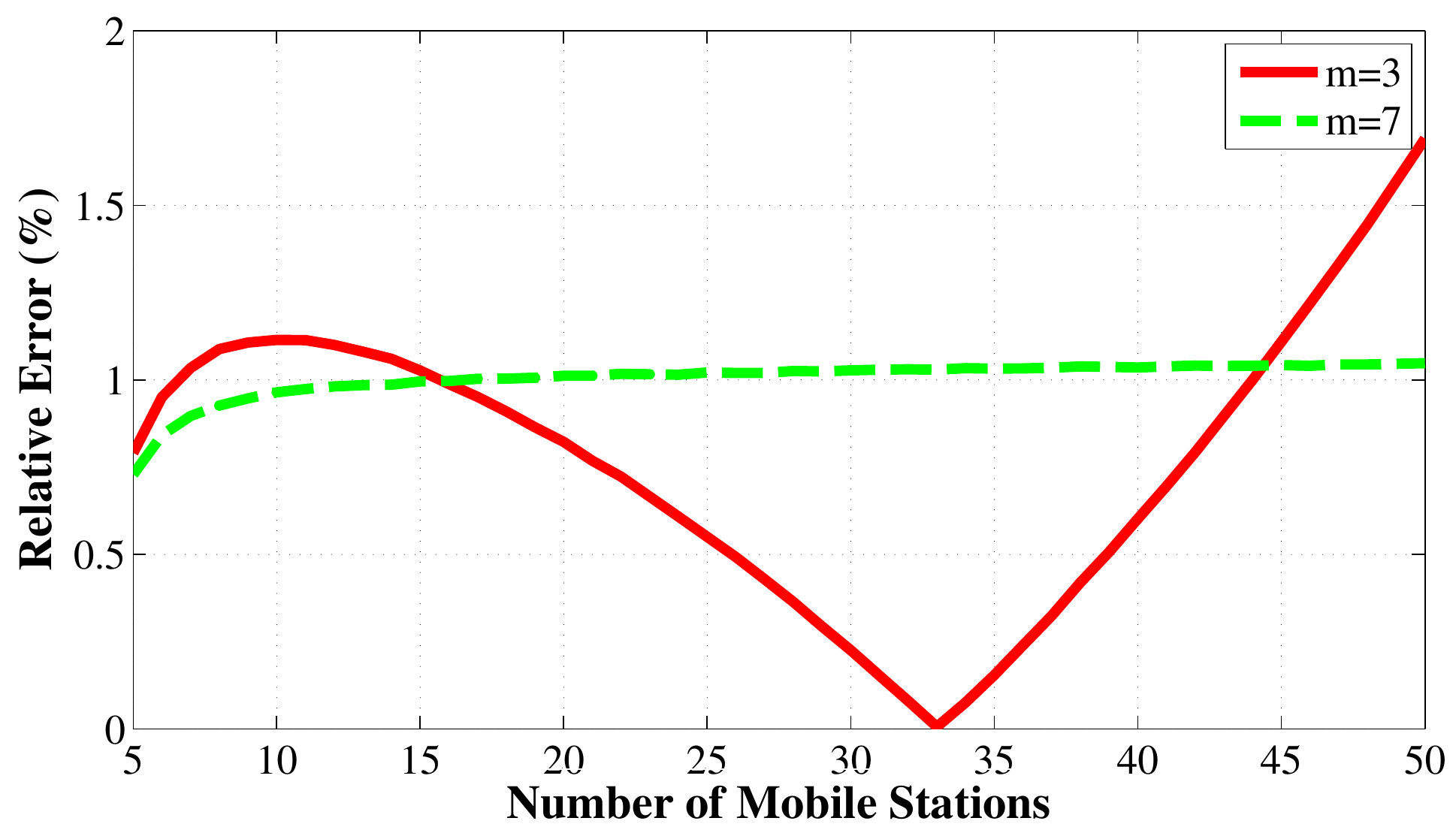}\\
	\caption{Relative error vs number of mobile stations.}
	\label{validation}
\end{center}
\end{figure}

In order to validate the analytical model, the proposed backoff strategy is simulated for various number of mobile stations. Saturation throughput is computed for 2 different maximum backoff stages ($m=3$ and $m=7$). Figure \ref{validation} illustrates the relative error. The difference between the analytical and the simulated model is negligeable and it is due to the solve function tolerance as well as the finite number of iteration considered in the simulation.

\subsection{System Performance}
\begin{figure}[tb]
\begin{center}
	\includegraphics[width=1\columnwidth]{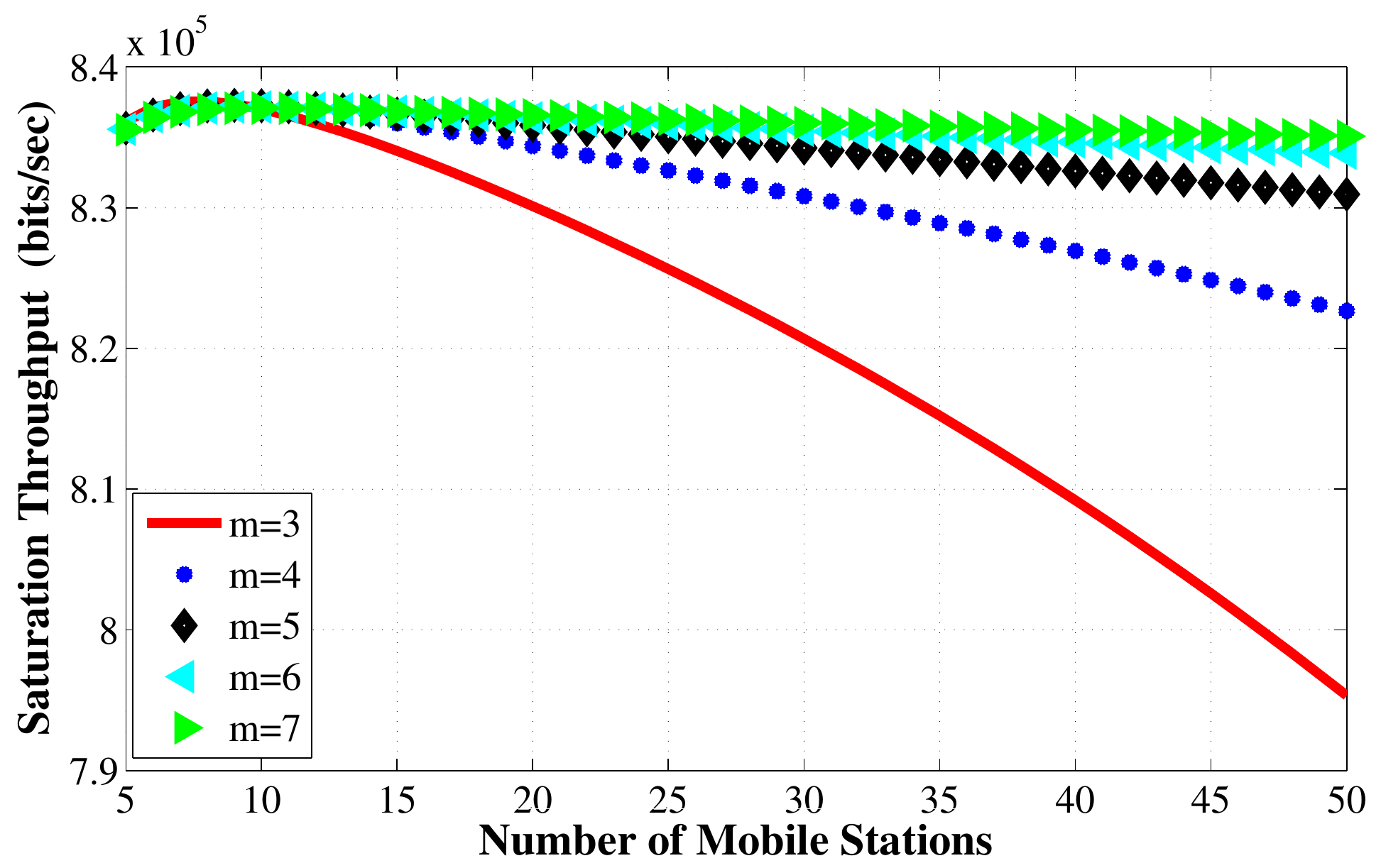}\\
	\caption{Saturation throughput for proposed strategy with RTS/CTS transmission.}
	\label{proposedrtsctsthroughput}
\end{center}
\end{figure}

\begin{figure}[h!]
\begin{center}
	\includegraphics[width=1\columnwidth]{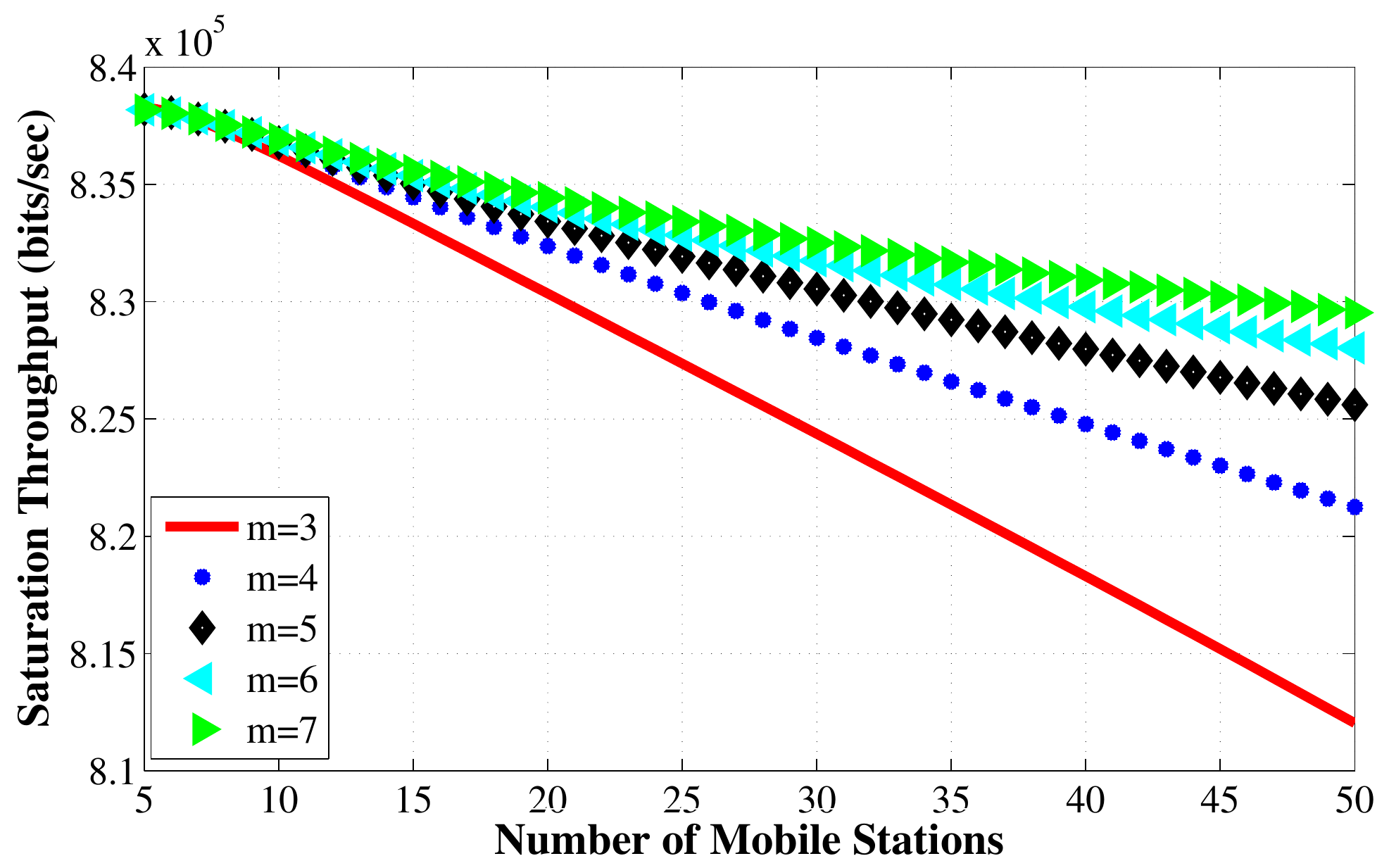}\\
	\caption{Saturation throughput for classical 802.11 FH with RTS/CTS transmission.}
	\label{classicrtsctsthroughput}
\end{center}
\end{figure}

In this Section, we study the performance of the proposed backoff strategy, so we compute the saturation throughput (bits/sec) vs the number of mobile stations for the RTS/CTS mode. RTS/CTS transmission mode is considered as it avoids the collision between the long data packets especially for high number of mobile stations. 
Figures \ref{proposedrtsctsthroughput} and \ref{classicrtsctsthroughput} show that the saturation throughput in the proposed strategy is better than the saturation throughput in the classical CSMA/CA protocol with 802.11 backoff strategy for RTS/CTS mechanism, and especially in the cases of large $CW_{max}$ (big $m$) independently from the number of mobile stations.

For example, in the proposed protocol and for $CW_{max}=511$ ($m=5$) we can achieve better saturation throughput than the 802.11 protocol with $CW_{max}=2047$ ($m=7$).

Numerical results show, as expected and due to lower probability of collision between transmitters, that the throughput increase when the number of states become higher. This fact is due to the distribution of all users within different backoff states, instead to be all located in the first state (bottleneck of classical 802.11 protocol).
Note that we don't take into consideration the retransmission limit and the maximum backoff stage as defined by the IEEE standard specification \cite{specs1}.
It should be mentioned that 802.11  strategy has better performance for large number of users and $m=3$ and it is due to the lack of spatial degree of liberty.
\subsection{Statistical Delay Study}
\begin{figure}[h!]
\begin{center}
	\includegraphics[width=1\columnwidth]{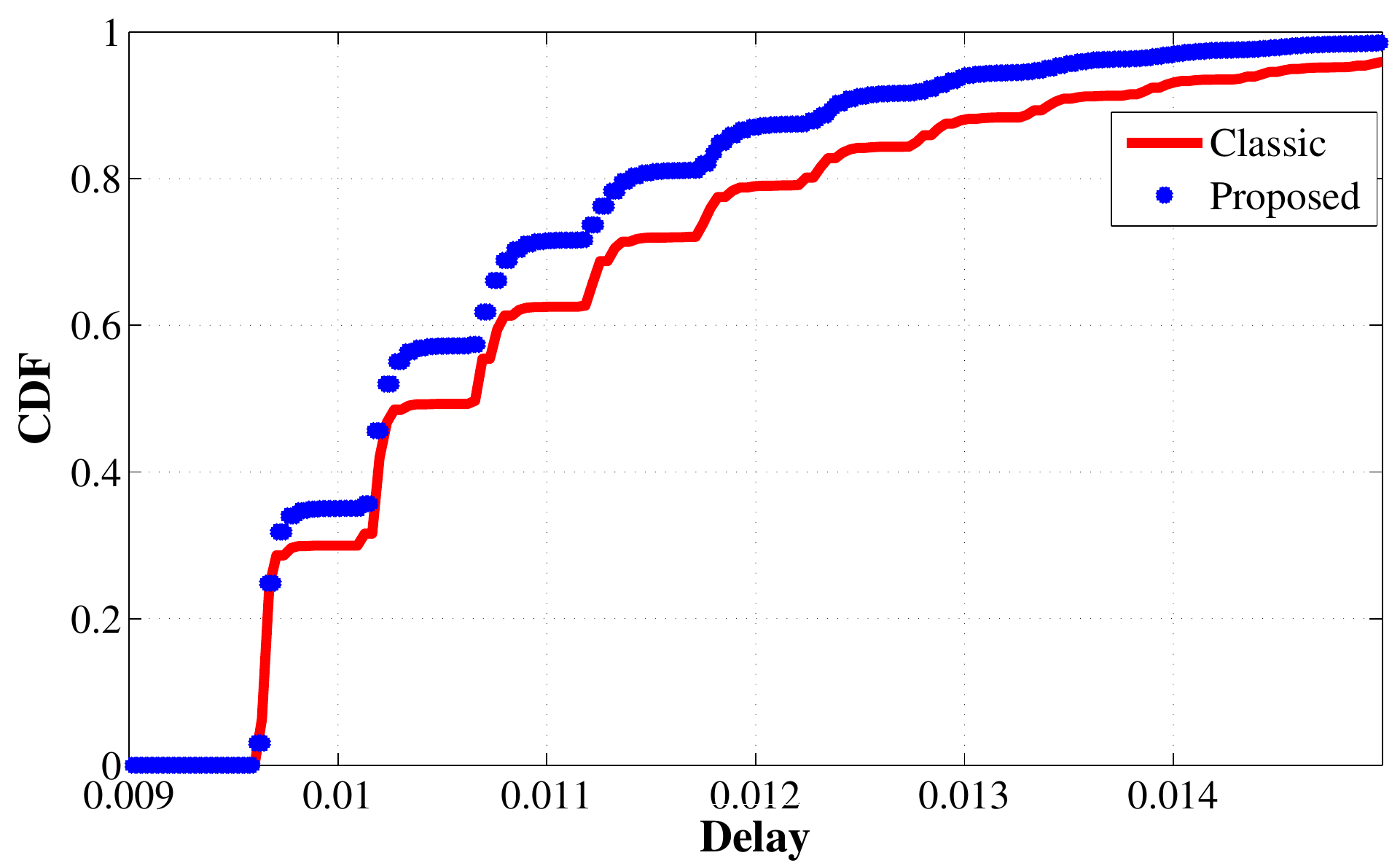}\\
	\caption{CDF of access delay for $m=3$ with 50 mobile stations. Delay is expressed in second.}
	\label{m3cdf}
\end{center}
\end{figure}

Many previous works \cite{novel,av1,av2,av3} evaluates the system performance in term of delay by computing or simulating the average access delay.
Since the average access delay isn't always a sufficient metric especially in VoIP applications, we go forward to simulate the cumulative density function (CDF) of the access delay. Figure \ref{m3cdf} represents the CDF of the access delay for $m=3$.
It's seen clearly from Figure \ref{m3cdf} that the delay of the proposed strategy is less than the classical one especially in charged mode (large number of mobile stations). It is due to the fact that users are distributed over all the stages instead to be located in the bottleneck ($m=1$). Also, the proposed backoff strategy is much more robust with high states number (big $m$) thanks to the offered spatial liberty. Tables \ref{m3table} and \ref{m7table} give different delay values for some CDF with an idea about the gain introduced by our strategy. 
For instance, 99\% of packets are transmitted with at most $17.6$ ms (resp $13.9$ ms) by the classical IEEE backoff while they are sent with at most $15.5$ ms (resp $12.5$ ms) by our proposed backoff strategy for $m=3$ (resp $m=7$).

\begin{table}[h!]
\caption{Delay ($ms$) and gain (\%) values in both backoff strategies for many CDF values with $m=3$}
\centering
\begin{tabular}{lccc}
\hline
CDF & Proposed Backoff ($ms$) & Classical Backoff ($ms$) & Gain (\%)\\
\hline
99\% & 15.5 & 17.6 & 11.93\\
98\% & 14.5 & 16.4 & 11.58\\
95\% & 13.3 & 14.6 & 8.90\\
90\% & 12.4 & 13.4 & 7.46\\
\hline
\end{tabular}
\label{m3table}
\end{table}

\begin{table}[tb]
\caption{Delay ($ms$) and gain (\%) values in both backoff strategies for many CDF values with $m=7$}
\centering
\begin{tabular}{lccc}
\hline
CDF & Proposed Backoff ($ms$) & Classical Backoff ($ms$) & Gain (\%)\\
\hline
99\% & 12.5 & 13.9 & 10.07\\
98\% & 12.0 & 13.2 & 9.09\\
95\% & 11.4 & 12.3 & 7.32\\
90\% & 10.9 & 11.5 & 5.22\\
\hline
\end{tabular}
\label{m7table}
\end{table}

%% file: conclusion.tex
In this paper, we proposed and developped an analytical model for a new backoff strategy for CSMA/CA-CTS/RTS protocol. We validated the analytical model by simulations and we proved that the saturation throughput performance and the statistical access delay are improved especially in loaded systems. This proposed strategy could be a good candidate to solve the bottleneck problem existing in the classical IEEE 802.11 backoff strategy.
Our model assumes a finite number of terminals and ideal channel conditions. The model is suited for both Basic and RTS/CTS access mechanisms.
To conclude, in this contribution we proposed a solution for the bottleneck problem, but still as future work to discuss the improvement relative to adaptive minimum contention window and to find an original manner to deal with different users bitrates.

%% file: acknowledgment.tex
Authors wish to thank Prof. Jean-Marie Gorce and Dr. Nikolai Lebedev for their discussions and important remarks given following their intensive review for this work.